\documentclass[prstab,amsmath,amssymb,reprint,longbibliography]{revtex4-1}
\usepackage[utf8]{inputenc}

\usepackage{color,graphicx,hyperref,siunitx}
\usepackage[caption=false]{subfig}

\usepackage{tikz}
\definecolor{dblu}{HTML}{000077}
\hypersetup{colorlinks=true, linkcolor=dblu, citecolor=dblu, filecolor=dblu, urlcolor=dblu}
\definecolor{lime}{HTML}{A6CE39}
\DeclareRobustCommand{\orcidicon}{
    \hspace{-4mm}
	\begin{tikzpicture}
	\draw[lime, fill=lime] (0,0) 
	circle [radius=0.16] 
	node[white] {{\fontfamily{qag}\selectfont \tiny ID}};
	\draw[white, fill=white] (-0.0625,0.095) 
	circle [radius=0.007];
	\end{tikzpicture}
}

\newcommand{\avg}[1]{\left\langle #1 \right\rangle}
\newcommand{\tod}{\mathrm d}
\newcommand{\eu}[1]{\,\mathrm e^{#1}}
\newcommand{\im}{\mathrm i}

\begin{document}

\title{The Mathieu unit cell as a template for low emittance lattices}

\thanks{submitted to Phys. Rev. Accel. Beams}
\date{February 24, 2021}
\author{B. Riemann\href{https://orcid.org/0000-0002-5102-9546}{\orcidicon}}
\affiliation{Paul Scherrer Institut, CH-5232 Villigen PSI, Switzerland}

\begin{abstract}
The multi-bend achromat (MBA), which often serves as a building block for modern low-emittance storage rings, is composed of a repetition of unit cells with optimized optical functions for low emittance in the achromat center, as well as end cells for dispersion and optics matching to insertion devices.

In this work, we describe the simplest stable class of unit cells that are based on a longitudinal Fourier expansion, transforming Hill equations to Mathieu equations. The resulting cell class exhibits continuously changing dipolar and quadrupolar moments along the beam path. Although this elementary model is defined by only three parameters, it captures a significant amount of notions that are applied in the design of MBAs. This is especially interesting as Mathieu cells can be viewed as an elementary extension of Christofilos' original model of alternating-gradient focusing, while their sinusoidal bending and focusing functions lend themselves to future applications in undulator-like structures.

Mathieu cells can be used to estimate the range of reasonable cell tunes and put an emphasis on the combination of longitudinal gradient bending and reverse bending, as well as on strong horizontal focusing to reach emittances lower than the classic theoretical minimum emittance cell. Furthermore, the lowest emittances in this model are accompanied by small absolute momentum compaction factors.
\end{abstract}

\maketitle

\section{Introduction}

For practical reasons, the evolution of lattices for low-emittance synchrotron storage rings, including the double-bend \cite{chasman-green}, triple-bend  \cite{vignola} and quadruple-bend achromats~\cite{einfeld-plesko}, is mostly based on modeling them with discrete elements representing accelerator magnets. This also includes computations for the emittance minimum of a given periodic cell \cite{sommer,teng-tme}.

Recent developments in the field of multi-bend achromats (MBAs) have shown that longitudinal gradients in magnet strength can significantly decrease emittance~(e.g.~\cite{nagaoka-wrulich,gradientbend,clic-trapz}), and that reverse bends~\cite{delahaye} (see also Veksler's suggestion in \cite{courant-snakes}) are necessary to fully exploit these longitudinal gradients~\cite{antibend,lgb-rb}. It has also been known for a long time that combined-function magnets can help to decrease horizontal emittance by manipulating damping partitions (see e.g.~\cite{vignola,einfeld-plesko}).

These facts can inspire to model the focusing and bending functions of the periodic lattice structure (i.e., the unit cell) directly, by a set of basis functions that are periodic in cell length,
instead of using distinct elements to represent magnets. In principle, the type of basis function can be selected in an arbitrary manner. E.g., in~\cite{numopt-tmePub} step-like basis functions are used and truncated at a high order, and a particle-swarm based optimization is applied in the resulting high-dimensional parameter space.

The choice of sinusoidal basis functions is motivated in Sec.~\ref{sec:longHarm} --  
in essence, higher harmonics of the unit cell require stronger magnet pole-tip fields than lower harmonics, which is especially important for miniaturized magnet arrangements, where the lowest harmonics will dominate. This statement can be related to the common treatment of undulators, which usually starts with a description of the lowest harmonics (e.g.~\cite{wille}).

The focusing functions for Mathieu cells, which we introduce in this work, contain the lowest possible order of such basis functions that yield stable solutions and are discussed in Sec.~\ref{sec:mathieu2d}. It is interesting to note that these sinusoidal focusing forces are also the starting point for Christofilos' description of alternating-gradient focusing \cite{christofilos}. However, his derivations focus on qualitative aspects of the motion, and not on solving the underlying differential equations -- these are Mathieu equations.

Afterwards, bending functions are included in Sec.~\ref{sec:bending}. The resulting parameter space is three-dimensional and can be explored without difficulty. For the resulting cells, synchrotron integrals, emittance and momentum compaction can be computed, and example solutions are studied.

The scaling laws for unit cells are investigated in Sec.~\ref{sec:sextupoles} with an emphasis on the `chromaticity wall' and selecting the optimal cell length. A new objective function for the emittance of an arc with optimally scaled cell length is obtained, including constraints on applicable sextupole field strength. After further approximating the applicable pole-tip fields of magnets for a specific example tune, an example cell is constructed using parameters of the SLS~2.0 storage ring in Sec.~\ref{sec:sls2}.

\section{Longitudinal harmonics\label{sec:longHarm}}

Consider the magnetic field on a cylinder with variable radius $r$, and the beam path leading through its axis. For simplicity, we neglect the curvature of the path, although the argument naturally extends to that case. In a current-free region, a scalar potential defining the magnetic field $\vec B = -\nabla \Psi$ obeys the Laplace equation $\nabla^2 \Psi = 0$~\cite{jackson}. In the aforementioned periodic cell, this potential can be expressed as linear combination of basis functions
\begin{align}
    \label{eq:V}
    \tilde \Psi_{n,p} = D_{|n|}(\bar k_p, r) \, \eu{\im n \phi} \eu{\im \bar k_p z}
\end{align}
for integers $p$ and $n$, and with the definition of
\begin{align}
    D_n(\bar k_p, r) &= 2^n I_n(\bar k_p r) \Big/ \bar k_p^n,
\end{align}
where $I_n$ is the modified Bessel function of the first kind and order $n$ (see Appendix~\ref{app:propD}). Defining the period of the cell to be $L$, one obtains $\bar k_p = 2 \pi p / L$. 

When selecting a longitudinal harmonic with positive $n$, the radial field component at radius $r$ is given as (Appendix~\ref{app:propD})
\begin{align}
    B_r \propto \frac{\tod D_n}{\tod r} \propto r^{n-1}
    \left(1 + \bar k_p^2 \frac{n+2}{4 n (n+1)} r^2 + \dots \right).
\end{align}
For $p=0$ this reduces to the commonly known behavior $B_r \propto r^{n-1}$. However, the higher the longitudinal harmonic $|p|$ to be considered, the more difficult an application of the desired on-axis multipolar fields will become.

Therefore, lower longitudinal harmonics of multipolar fields are preferable to higher harmonics.
Further assuming the unit cell to possess symmetry planes, we can select cosine functions $\cos(\bar k_p s)$ as basis functions with increasing positive order $p \leq P$.

\subsection{Biplanar stability}

The next task is finding the lowest maximum order $P$ for which stable particle motion could be achieved. The transverse linear motion of a charged particle with design energy in a decoupled accelerator lattice without bending magnets can be described using Hill differential equations~\cite{hill,courant-snyder}
\begin{align}
   \frac{\tod^2}{\tod s^2} x(s) + \kappa(s)\;x(s) &= 0,\nonumber\\
   \frac{\tod^2}{\tod s^2} y(s) - \kappa(s)\;y(s) &= 0.\label{eq:hill}
\end{align}
Note that bending magnets (Sec.~\ref{sec:bending}), chromatic effects (Sec.~\ref{sec:chroma}, Sec.~\ref{sec:sextupoles}) and fringe effects (Sec.~\ref{sec:fringe}) are discussed in later sections.

Assuming $\kappa(s)$ to be constructed of basis functions $\cos(\bar k_p s)$, the most elementary case to consider is $P=0$ because then $\kappa=\text{const}$. As the sign of $\kappa$ is different for the horizontal and vertical plane, bounded motion can only be achieved in one of them, and stable particle motion is impossible.

On the other hand, as we will see, the case $P=1$ already allows for stable motion. The resulting parameter space is low-dimensional, and thus lends itself to plain exploration.
We first investigate such a model without bending and thus without dispersion. An additional parameter for bending is then included, and synchrotron radiation integrals (including damping partition, emittance, momentum compaction) are computable.

\section{Mathieu equations in 2d\label{sec:mathieu2d}}

To simplify the following calculations, we consider a normalized cell with the dimensionless length $\pi$.
The normalized longitudinal cell coordinate $u$ is linked to the standard cell coordinate  via $s=L u / \pi$ (see Appendix~\ref{app:fxArc}).

Still considering the aforementioned focusing function for the case $P=1$, we obtain
\begin{equation}
\label{eq:xyHom}
\begin{aligned}
  \frac{\tod^2}{\tod u^2} x(u) + k(u) \;x(u) &= 0,\\
  \frac{\tod^2}{\tod u^2} y(u) - k(u)\;y(u) &= 0 
\end{aligned}
\end{equation}
with a cell-normalized focusing strength
\begin{align}
  k(u) = k_0 - 2 k_1 \cos(2 u),
\end{align}
where the factor $-2$ was selected arbitrarily for alignment with standard notation. The equations of motion are now Mathieu equations, both depending on the same set of parameters $k_0,k_1$.

We analyze the horizontal motion based on Floquet solutions, mainly following the approach outlined in~\cite{characteristicExponents}. These can be written in the normal form~\cite{chicone-ode}
\begin{align}
  \label{eq:xSol}
  X(u) = \eu{\im 2 \nu_\mathrm x u} f(u),
\end{align}
where $2 \nu_\mathrm x$ is the characteristic exponent and $\nu_\mathrm x$ is the horizontal cell tune, i.e., the betatron phase advance in a cell divided by $2\pi$. 
We express the $\pi$-periodic function
\begin{align}
  f(u) &= \sum_{q=-Q}^Q f_q \eu{\im 2 q u}
\end{align}
as a truncated Fourier series with the highest harmonic being $Q$. For the following calculations, $Q=50$ is sufficient. Then using the $f_q$ as components of a vector $\vec f$, we can write Eq.~\eqref{eq:xyHom} as
\begin{align}
  \label{eq:sysM}
  0 &= \mathbf M(\nu_\mathrm x) \; \vec f,
\end{align}
where $\mathbf M(\nu_\mathrm x)$ is a (truncated) tridiagonal matrix of size \\$(2Q+1) \times (2Q+1)$ with entries
\begin{align}
  M_{q,q} = 1 &,\quad M_{q,q-1} = M_{q,q+1} = \frac{ k_1 }{4 (q + \nu_\mathrm x)^2 - k_0}.
\end{align}

\begin{figure}[!b]
    \centering
    \includegraphics[scale=0.9]{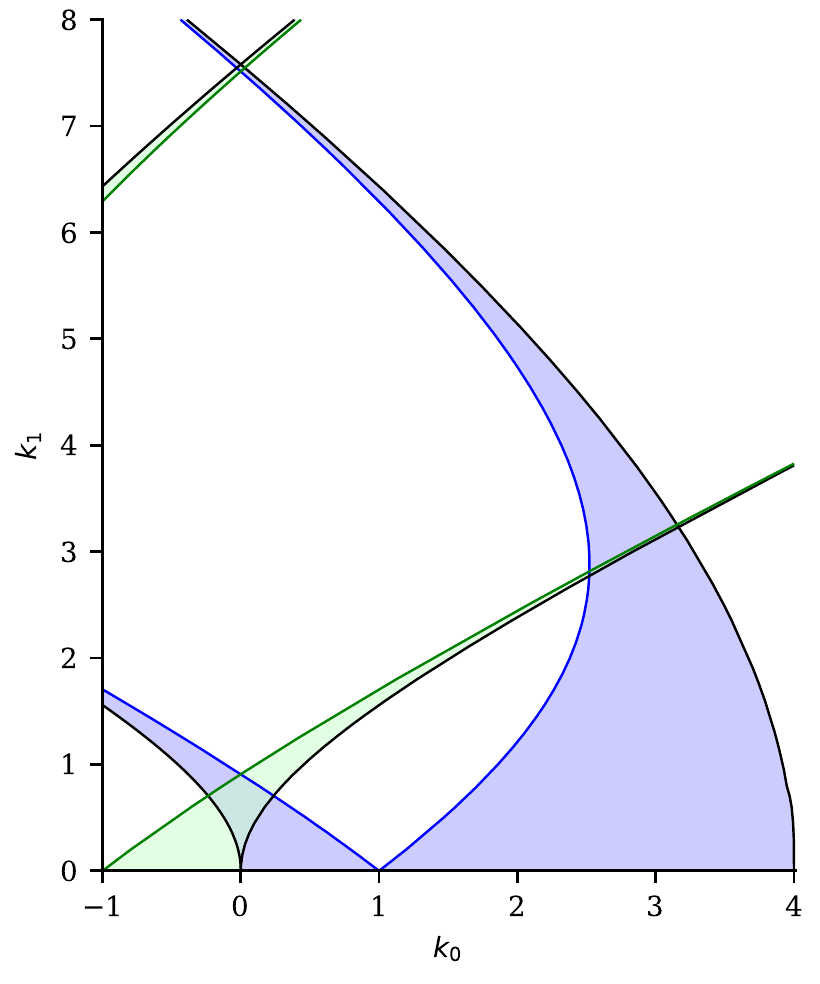}
    \caption{Stability diagram in the $(k_0,k_1)$ plane. Blue-shaded regions are stable for horizontal motion, with the blue line indicating $\nu_\mathrm x=0.5$. Green-shaded regions are stable for vertical motion, with the green line indicating $\nu_\mathrm y=0.5$. The region overlaps are stability islands.\label{fig:islands}}
\end{figure}

To solve this system, we require prior knowledge of $\nu_\mathrm x$. This can be achieved using the determinant of $\mathbf M(\nu_\mathrm x=0)$ and the Whittaker-Hill formula~\cite{characteristicExponents,mathieuApprox}
\begin{align}
  \sin^2 (\pi \nu_\mathrm x) &= C(k_0,k_1) \label{eq:sinCk}
\end{align}
with the definition
\begin{align}
  C(k_0,k_1) &= \det \mathbf M(\nu_\mathrm x=0) \cdot \sin^2 (\pi \sqrt{k_0}/2).
\end{align}
It is apparent from Eq.~\eqref{eq:sinCk} that periodic solutions only exist for
\begin{align}
  0 < C(k_0,k_1) < 1, 
\end{align}
leading to limited regions in $(k_0,k_1)$ space where horizontally stable motion occurs.
Furthermore, it follows from Eq.~\eqref{eq:xyHom} that stability of vertical motion is equivalent to that of horizontal motion when mirroring the $(k_0,k_1)$ regions at the origin. The intersection of stability regions for both planes leads to islands of stability for transverse motion (see Fig.~\ref{fig:islands}, cf.~\cite[Fig.~5]{mathieuApprox}). 

\begin{figure}[!t]
    \centering
    \includegraphics{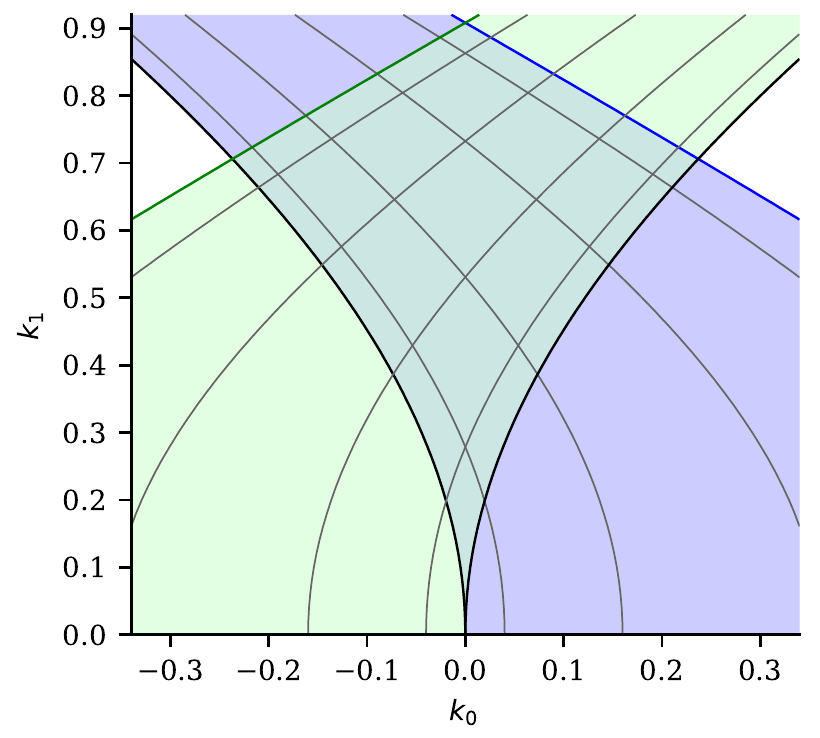}
    \caption{Stability diagram for the neck-tie island. Shading and colored lines like in Fig.~\ref{fig:islands}. Gray lines are isolines of $\nu_\mathrm x$ or $\nu_\mathrm y$ in steps of $0.1$.\label{fig:necktie}}
\end{figure}

The islands differ significantly in the maximal focusing strength that needs to be applied.
The only stable solutions with reasonable $\max |k(u)| \leq 2$ all occur in a single stability island. This `neck-tie' island, named here in analogy to the corresponding diagram for the FODO lattice~\cite{wiedemann} is shown in Fig.~\ref{fig:necktie} in more detail.
We conclude that reasonable cell designs require $(k_0,k_1)$ in this island, which has cell tunes $\nu_\mathrm x,\nu_\mathrm y < 1/2$.

\subsection{Tune map for chromaticity\label{sec:chroma}}

As there exists a bijective mapping of stable-motion quadrupole configurations to tunes $(k_0,k_1) \leftrightarrow (\nu_\mathrm x,\nu_\mathrm y)$, we are able to study the properties of Mathieu unit cells directly in tune space.

Given $\nu_\mathrm x$, one may solve Eq.~\eqref{eq:sysM} for $f(u)$. Optical functions are computed from $X(u)$ in Eq.~\eqref{eq:xSol} as (see e.g.~\cite{riemann-phd})
\begin{align}
  \label{eq:beta}
  \tilde\beta_\mathrm x(u) &= X(u) X^*(u) \Big/ I_\mathrm x = f(u) f^*(u) \Big/ I_\mathrm x 
\end{align}
with
\begin{align}
  I_\mathrm x &= \Im \big\lbrace X^*(u) X^\prime(u) \big\rbrace 
  = \Im \big\lbrace f^*(u) f^\prime(u) \big\rbrace,
\end{align}
and they can be used to compute the linear chromaticity with the horizontal and vertical optical functions $\tilde\beta_\mathrm x,\tilde\beta_\mathrm y$ by (cf.~\cite{wille})
\begin{equation}
\label{eq:chroma}
\begin{aligned}
    4 \pi \xi_\mathrm x &= - \int_0^\pi \tilde\beta_\mathrm x(u) k(u) \,\tod u, \\
    4 \pi \xi_\mathrm y &= \phantom+ \int_0^\pi \tilde\beta_\mathrm y(u) k(u) \,\tod u.
\end{aligned}
\end{equation}
Following from the aforementioned symmetry of vertical and horizontal motion in $(k_0,k_1)$ space, we obtain the vertical chromaticity for a given tune as 
\begin{align}
    \xi_\mathrm y(\nu_\mathrm x,\nu_\mathrm y) = \xi_\mathrm x(\nu_\mathrm y,\nu_\mathrm x).
\end{align}

\begin{figure}[!b]
    \centering
    \includegraphics{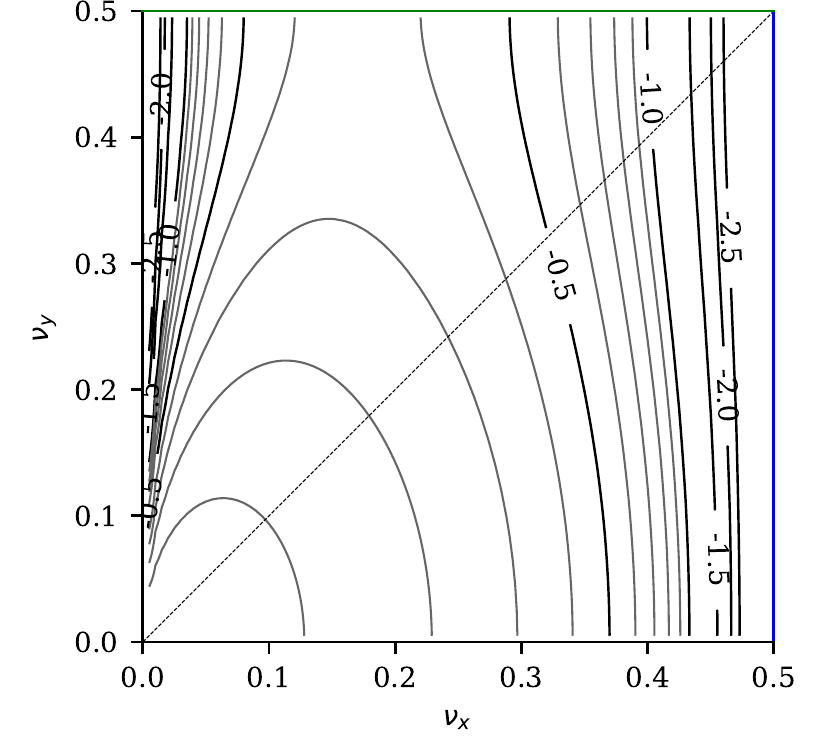}
    \caption{Tune map for horizontal linear chromaticity. The borders of the neck-tie island are shown as colored edges of the plot, corresponding to the lines in Fig.~\ref{fig:necktie}.\label{fig:chroma}}
\end{figure}

The results of the linear chromaticity computation are shown in Fig.~\ref{fig:chroma}. In the usable regions of the tune map, i.e., considering stop-bands around the half-integer resonances, we obtain negative chromaticities $\xi_{\mathrm x,\mathrm y} > -2.5$.

Note that the general dependency of cell tune on particle energy -- without effects by higher-order multipoles yet to be introduced -- can be obtained by scaling the $(k_0,k_1)$ vector corresponding to a given tune in the neck-tie diagram in Fig.~\ref{fig:necktie}.

\section{Bending and emittance\label{sec:bending}}

The next task is to include bending into the unit cell. We assume that the curvature is sufficiently small so that we can neglect the effect of weak focusing on $k(s)$ in Eq.~\eqref{eq:xyHom}. Following the same line of reasoning we also neglect edge focusing.

The average curvature $\avg{1/\rho}$ in a cell is defined by the arc geometry of the storage ring and the bending of end-cells.
When assuming the curvature to contain low-order longitudinal harmonics in the same manner as the focusing strengths $(P=1)$, we can parameterize
\begin{align}
   1/ \rho(u) = \avg{1/\rho} b(u)
\end{align}
with a cell-normalized dipole strength
\begin{align}
  b(u) =  1 - 2 b_1 \cos(2 u).
\end{align}
Note that $b(u)$ shares the same mirror symmetry around $u=0$ and $u=\pi/2$ as $k(s)$, as this is the most elementary approach.

Also, an upper limit on $|b|$, and thus $|b_1|$, exists given by achievable dipole field strength independent of cell length, as
\begin{align}
  \label{eq:maxB}
  \max|b| = \frac{\max|B|}{B_\text c}.
\end{align}
Here we introduced the characteristic magnetic field density
\begin{align}
    \label{eq:charB}
    B_\text c = (B \rho) \avg{1 / \rho}
\end{align}
depending on the beam rigidity $(B \rho)$.

Normalizing with the average curvature, the inhomogeneous Hill equation for linear dispersion $\eta(s)$~\cite{wille, wiedemann} can be rewritten as (see Appendix~\ref{app:fxArc})
\begin{align}
  \label{eq:dispersion}
  \frac{\tod^2}{\tod u^2} \tilde\eta(u) + k(u)\;\tilde\eta(u) &= b(u)
\end{align}  
with 
\begin{align}
  \tilde\eta(u) &= \left( \frac{\pi}L \right)^2 \frac{\eta(u)}{\avg{1/\rho}}.
\end{align}

We recognize that the solutions $\tilde\eta(u)$ of Eq.~\eqref{eq:dispersion} are additive in $b(u)$. 
Let $\tilde\eta^{(0)}(u)$ be the solution for $b(u)=1$ and let $\tilde\eta^{(1)}(u)$ be the solution for $b(u)=-2\cos(2u)$. 
Then the general solution is linear in $b_1$, as
\begin{align}
  \label{eq:linear_eta}
  \tilde\eta(u) &= \tilde\eta^{(0)}(u) + b_1\tilde\eta^{(1)}(u).
\end{align}
The driving term $b(u)$ requires $\tilde \eta(u)$ to be periodic in $\pi$,
\begin{align}
  \tilde \eta(u) = v_0 + 2 \sum_{q=1}^Q v_q \cos(2 q u) &= \sum_{q=-Q}^Q v_q \eu{2 \im q u},
\end{align}
so that Eq.~\eqref{eq:dispersion} reduces to the solvable linear equation system
\begin{align}
 \mathbf M(\nu_\mathrm x=0) \; \vec v &= \vec c
\end{align}
with $c_0=1/k_0$, $c_1 = c_{-1} = b_1 / (4 - k_0)$ and all other components of $\vec c$ being zero. The solution $\tilde \eta(u)$ can then be constructed using $\vec v$.

\begin{figure*}[t]
    \includegraphics{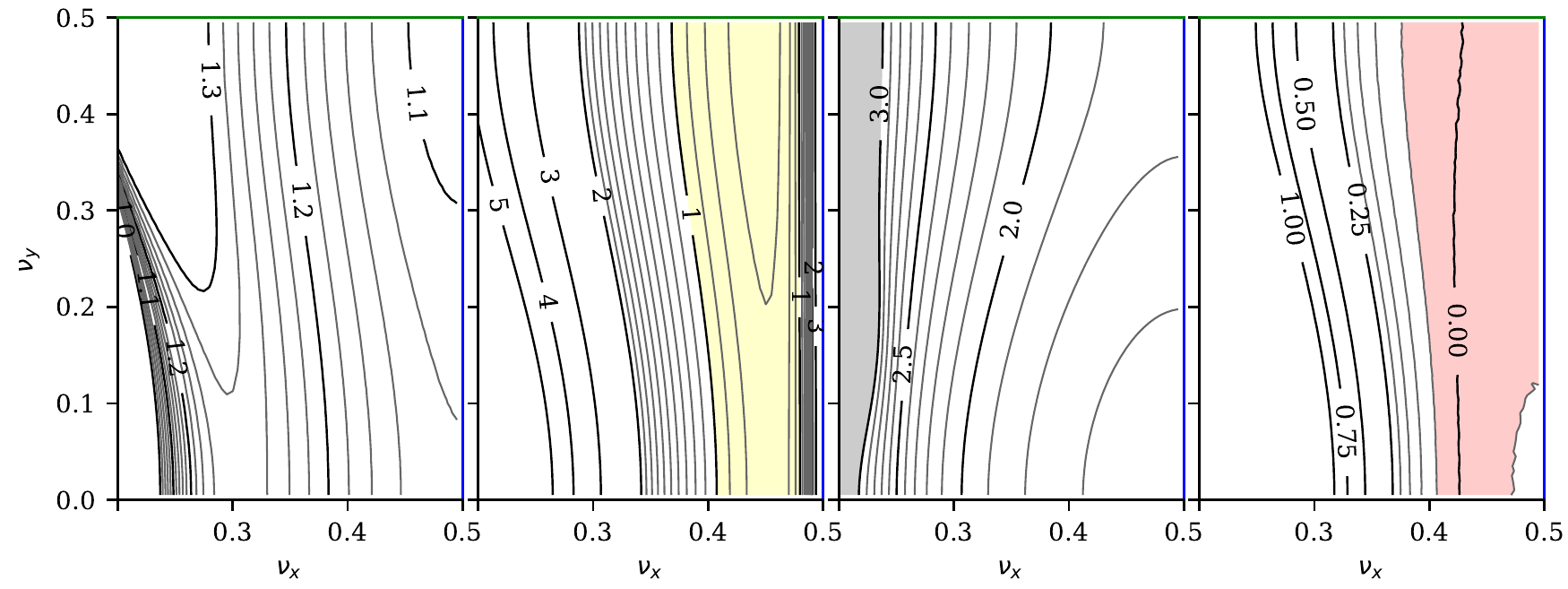}
    \caption{From left to right: (1) optimized $b_1$ for minimal TME emittance ratio $F$ for $b_1 < 1.3$, (2) resulting $F$ up to 5, with sub-TME region in yellow, (3) horizontal damping partition $J_\mathrm x$, invalid region in gray,
    (4) momentum compaction integral $I_1$ up to \num{1}, with the red area indicating $|I_1| < \num{0.05}$.\label{fig:F}}
\end{figure*}

\subsection{Synchrotron integrals}

Having introduced bending and dispersion, knowledge of linear momentum compaction can be obtained, which is proportional to the synchrotron integral~\cite{helm-integrals,sands} for the unit cell
\begin{align}
  I_1 = \int_0^\pi b(u)\; \tilde\eta(u) \,\tod u.
\end{align} 
To gain some insight into the behavior of $I_1$, we insert Eq.~\eqref{eq:linear_eta}
and obtain
\begin{align}
  \label{eq:quad_i1}
  I_1(b_1) &= 
  \int_0^\pi \tilde\eta^{(0)}(u)\,\tod u \nonumber\\ 
  &+ b_1 \left[ \int_0^\pi \tilde\eta^{(1)}(u)\,\tod u - 2 \int_0^\pi \tilde\eta^{(0)}(u) \cos(2u)\,\tod u\right] \nonumber\\
  &- b_1^2 \int_0^\pi \tilde\eta^{(1)}(u) [1 + \cos(4u)]\,\tod u.
\end{align}
By its definition preceding Eq.~\eqref{eq:linear_eta} and due to symmetry conditions, 
\begin{align}
 \int_0^\pi\tilde\eta^{(1)} \,\tod u = 0.
\end{align}
Although $\tilde\eta(u)$ is the solution of a parametric oscillator, we may expect it to mainly oscillate at the driving frequency $\cos(2u)$, making the last coefficient in Eq.~\eqref{eq:quad_i1} small.

We proceed by computing radiation properties for the normalized cell. The synchrotron integrals related to radiation loss and damping partitions are~\cite{helm-integrals,sands}
\begin{align}
  I_2 &= \int_0^\pi b(u)^2 \,\tod u = \pi (1 + 2 b_1^2)
\end{align}
and  
\begin{align}
  I_4 &\approx 2 \int_0^\pi b(u) k(u) \tilde\eta(u) \,\tod u.
\end{align}
The expression used for $I_4$ is an approximation in which, in consistence with our assumption, the contribution of weak focusing has been omitted. In full analogy to $I_1$ and substituting $\tilde\eta^{(\cdot)} \rightarrow k \tilde\eta^{(\cdot)}$, we find that $I_4(b_1)$ is also a quadratic function of $b_1$. 

In order for a flat lattice to allow damping in all dimensions, the horizontal damping partition
\begin{align}
  J_\mathrm x = 1 - \frac{I_4}{I_2}
\end{align}
must fulfill $0 < J_\mathrm x < 3$~\cite{helm-integrals,sands}.

In low-emittance rings, $J_\mathrm x > 1$ is favored~\cite{hmba16} as the effects of quantum excitation are then shifted from the transverse into the longitudinal plane.

The dispersion action $\mathcal H(s)$ occuring in the quantum excitation integral 
\begin{align}
  I_5 = \int_0^\pi \mathcal H(u)\;|b(u)^3| \,\tod u
\end{align}
can be computed using the Floquet solution as
\begin{align}
  \mathcal H(u) &= \tilde\gamma \tilde\eta^2 + 2 \tilde\alpha \tilde\eta \tilde\eta' + \tilde\beta \tilde\eta^{\prime 2} \\
  &= |X^\prime(u)\;\tilde\eta(u) - X(u)\;\tilde\eta^\prime(u)|^2 \Big/ I_\mathrm x.\nonumber
\end{align}
One can then obtain the emittance $\epsilon \propto I_5 / (I_2 J_\mathrm x)$. However, we are interested in the emittance relative to that of a normalized theoretical minimum emittance (TME) cell~\cite{teng-tme},
\begin{align}
  \label{eq:F}
  F(\nu_\mathrm x,\nu_\mathrm y,b_1) =  \frac{I_5}{I_2 J_\mathrm x} \Big/ \left( \frac{I_5}{I_2} \right)_\mathrm{TME}
  = \frac{12\sqrt{15}}{\pi^3} \frac{I_5}{I_2 J_\mathrm x},
\end{align}
as it is independent of cell length.

\subsection{Results}

We can now search for the optimal $b_1$ parameter to reach minimum emittance ratio $F$ for a given tune $(\nu_\mathrm x,\nu_\mathrm y)$; the results are shown in Fig.~\ref{fig:F}. Sub-TME emittances are reached for $0.4 < \nu_\mathrm x < 0.5$, with a minimal $F < 0.7$. We see that, in this band, increasing $\nu_\mathrm y$ only has slight effects -- increasing $J_\mathrm x$ and decreasing $F$. Damping partitions for the sub-TME region are in a feasible interval $J_\mathrm x \in [1.5,2.5]$.

The region with small absolute momentum compaction in Fig.~\ref{fig:F} has a similar location and shape as that of sub-TME emittance -- this is consistent with the general observation that low-emittance lattices require small absolute momentum compaction.

\begin{figure}[!htbp]
    \includegraphics{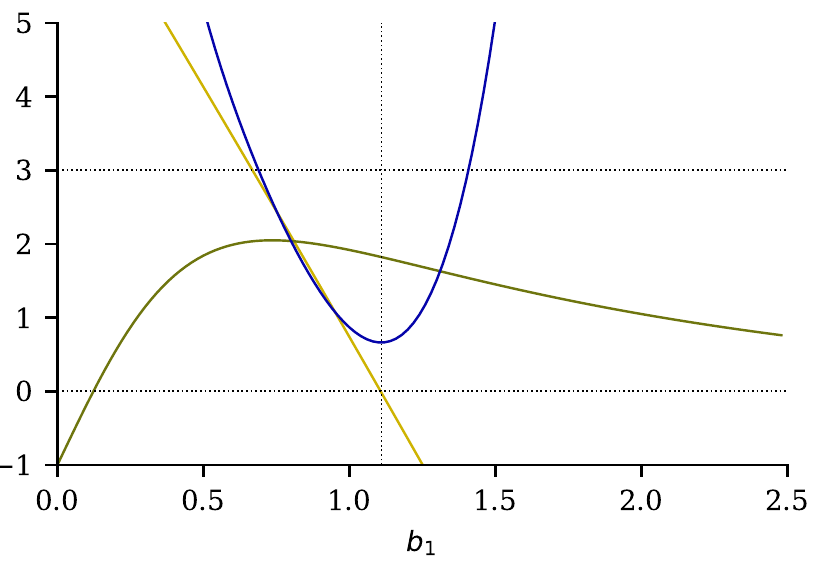}
    \caption{TME emittance ratio $F$ (blue), horizontal damping partition $J_\mathrm x$ (green) and momentum compaction integral $I_1$ (yellow) in dependence of the dipole coefficient $b_1$. The limits of $J_\mathrm x$ for stable motion, as well as the $b_1$ value used in the example solution, are denoted with dashed lines.\label{fig:b1scanA-int}}
\end{figure}

To further investigate the influence of the parameter $b_1$, which is not visible in the projections in Fig.~\ref{fig:F}, figures of merit for an example tune $\nu_\mathrm x=0.45,\nu_\mathrm y=0.35$ and variable dipole coefficient are shown in Fig.~\ref{fig:b1scanA-int}. According to Eq.~\eqref{eq:quad_i1} we expect $I_1$ to be quadratic in $b_1$, with the quadratic coefficient almost vanishing -- we obtain a visibly linear dependency here. The location of $I_1=0$ and the location of the minimal $F$ again illustrate that low emittances and low momentum compaction are closely related.

\begin{figure}[!b]
    \centering
    \includegraphics{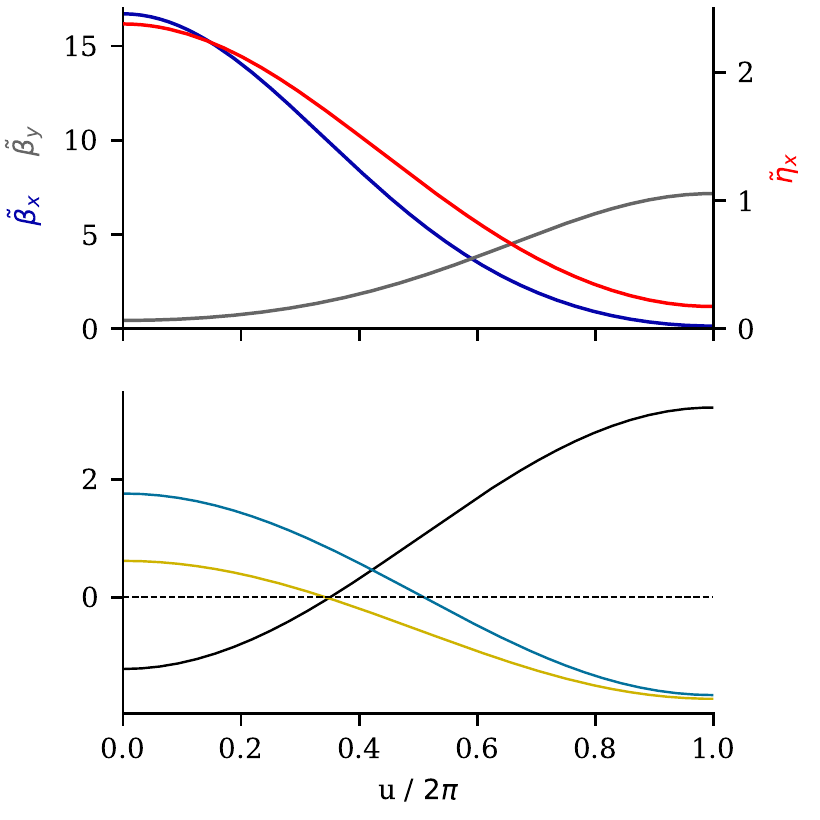}
    \caption{Top: optical functions for the example solution characterized in Table~\ref{tab:example}. Bottom: corresponding distribution of dipole (black), quadrupole (blue) and sextupole fields (yellow).\label{fig:example}}
\end{figure}

As the damping partition is in a usable range, the minimum emittance solution for this tune is feasible. The example solution parameters, figures of merit, and optical functions are shown in 
Fig.~\ref{fig:example} and Table~\ref{tab:example}. It can be seen that (1)~positive bending and defocusing quadrupole fields overlap, increasing $J_\mathrm x$~\cite{einfeld-plesko}, and that (2)~reverse bending occurs at the position of maximum dispersion~\cite{antibend}.

\begin{table}[!htbp]
    \caption{Example multipole parameters of a Mathieu cell and its figures of merit.\label{tab:example}}
    \begin{ruledtabular}
    \begin{tabular}{rrr}
    \textbf{Parameter} & & \textbf{Value} \\ \hline
    Dipole coefficient & $b_1$ & \num{-1.1100} \\    
    Quadrupole coefficient & $k_0$ & \num{0.04801} \\    
    & $k_1$ & \num{0.8554} \\ 
    Sextupole coefficient & $m_0$ & \num{-0.5550} \\
    & $m_1$ & \num{0.5855} \\ \hline
    horizontal cell tune & $\nu_\mathrm x$ & \num{0.4500} \\
    vertical cell tune & $\nu_\mathrm y$ & \num{0.3500} \\
    nat. horizontal cell chromaticity & $\xi_\mathrm x$ & \num{-1.8323} \\
    nat. vertical cell chromaticity & $\xi_\mathrm y$ & \num{-0.6716} \\ \hline
    horizontal damping partition & $J_\mathrm x$ & \num{1.8219} \\
    radiation integral & $I_1$ & \num{-0.0191} \\
    TME ratio & $F$ & \num{0.6617} \\
    arc emittance factor (sec.~\ref{sec:sextupoles}) & $G$ & \num{0.9963} \\
    \end{tabular}
    \end{ruledtabular}
\end{table}

\section{Sextupoles and chromaticity wall\label{sec:sextupoles}}

\begin{figure}[!b]
    \includegraphics{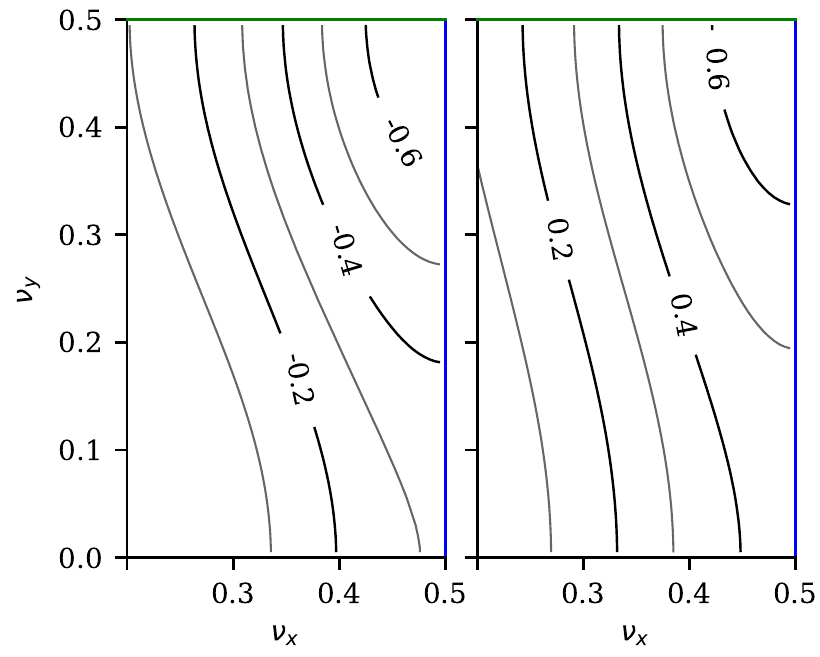}
    \caption{Values of $m_0$ (left) and $m_1$ (right) for optimized emittance ratio $F$ at a given cell tune.\label{fig:sextuVals}}
\end{figure}

To control linear chromaticity occurring according to Eq.~\eqref{eq:chroma}, the introduction of sextupolar fields
\begin{align}
    \label{eq:defMu}
    \mu(s) &= \frac 1{2 (B \rho)} \frac{d^2 B_y}{d x^2}(x=0,s),
\end{align}
is required. Full compensation leads to the condition \cite{wiedemann}
\begin{align}
    4 \pi \xi_\mathrm x &\stackrel != -\int \beta_x(s) \eta_x(s) \mu(s) \tod s, \nonumber\\
    4 \pi \xi_\mathrm y &\stackrel != \phantom+ \int \beta_y(s) \eta_x(s) \mu(s) \tod s. 
\end{align}
We define a cell-normalized sextupolar field function
\begin{align}
  m(u) &= m_0 + 2 m_1 \cos(2 u),
\end{align}
which includes the fundamental harmonic and shares the same mirror symmetry as $b(u)$ and $k(u)$, yielding the most elementary model,

For the unit cell with length $\pi$, this results (Appendix~\ref{app:fxArc}) in an equation system
\begin{align}
  \xi_\mathrm x = -\frac 12 \avg{\tilde\beta_x \tilde\eta m} &= -\frac{\avg{ \tilde\eta \tilde\beta_\mathrm x}}2 \; m_0 - \avg{ \tilde\eta \tilde\beta_\mathrm x \cos(2 u)} \; m_1, \nonumber\\
  \xi_\mathrm y = \phantom-\frac 12 \avg{\tilde\beta_y \tilde\eta m} &= \phantom-\frac{\avg{ \tilde\eta \tilde\beta_\mathrm y}}2 \; m_0 + \avg{ \tilde \eta \tilde \beta_\mathrm y \cos(2 u)} \; m_1, \label{eq:m1chroma}
\end{align}
which can be uniquely solved for $m_0, m_1$; $\avg{\cdot}$ denotes the average of the respective quantity over the cell length. The sextupole coefficients for $F$--optimized cells with given cell tunes are shown in Fig.~\ref{fig:sextuVals} and also included for the example in Table~\ref{tab:example}.

\subsection{Sextupole-limited arc emittance}

We want to find the cell length yielding the optimal emittance for a given limited sextupole strength $\max |m|$.
In an arc of constant average curvature $\avg{1 / \rho}$, the actual sextupole strength $\mu(u)$ scales relative to the sextupole strength of the normalized cell $m(u)$ as (see Appendix~\ref{app:fxArc})
\begin{align}
    \mu(u) = \left( \frac{\pi}L \right)^4 \frac{m(u)}{\avg{1/\rho}}. 
\end{align}
This disadvantageous dependency on cell length is sometimes referred to as `chromaticity wall' \cite{johan-ulels} and is a major limitation for shrinking unit cells. 

The optimal cell length can be obtained from the above equation as
\begin{align}
  \label{eq:opt-length}
  L = \pi \left( \frac{\max |m|}{\max |\mu| \avg{1/\rho}} \right)^{1/4}.
\end{align}
It is well known (e.g.~\cite{icfa-57}) that the emittance scales with the cube of bending angle per cell, and thus in our case $\propto L^3$. Reusing the definition of TME-normalized emittance $F$ in Eq.~\eqref{eq:F}, we find that the optimal emittance scales as
\begin{align}
    \epsilon &\propto F L^3 \propto G,
    &\text{with} \quad G &= F\; (\max |m|)^{3/4}.
\end{align}
We can use $G$ as an objective function for optimization, thus including sextupolar fields in a straightforward manner, to find an optimal value for $b_1$. 

Tune maps for figures of merit in which $b_1$ is selected to yield the optimal $G$ are shown in Fig.~\ref{fig:G}. We can observe that the characteristics for the emittance ratio $F$ and the damping partition $J_\mathrm x$ did not change significantly, although the tune-space region of low momentum compaction has reduced in size.

\begin{figure}[!tb]
    \includegraphics{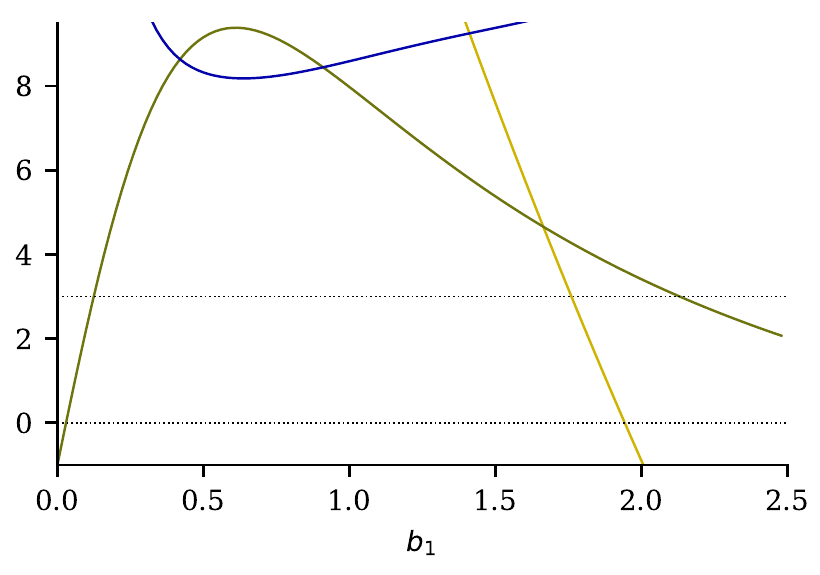}
    \caption{Figures of merit in dependence of the dipole coefficient $b_1$ (see Fig.~\ref{fig:b1scanA-int}) for $\nu_\mathrm x=\num{0.15}$, $\nu_\mathrm y=\num{0.35}$.}
    \label{fig:b1scanB-int}
\end{figure}

Furthermore, it is interesting that the two regions with $G \leq 1$ exist. One region has a significantly reduced horizontal focusing $\nu_\mathrm x < 0.2$. Unfortunately, the low $G$ values in this region are mainly influenced by a large and infeasible damping partition $J_\mathrm x > 3$ (see Fig.~\ref{fig:b1scanB-int}). 

\begin{figure}[!tb]
    \centering
    \includegraphics{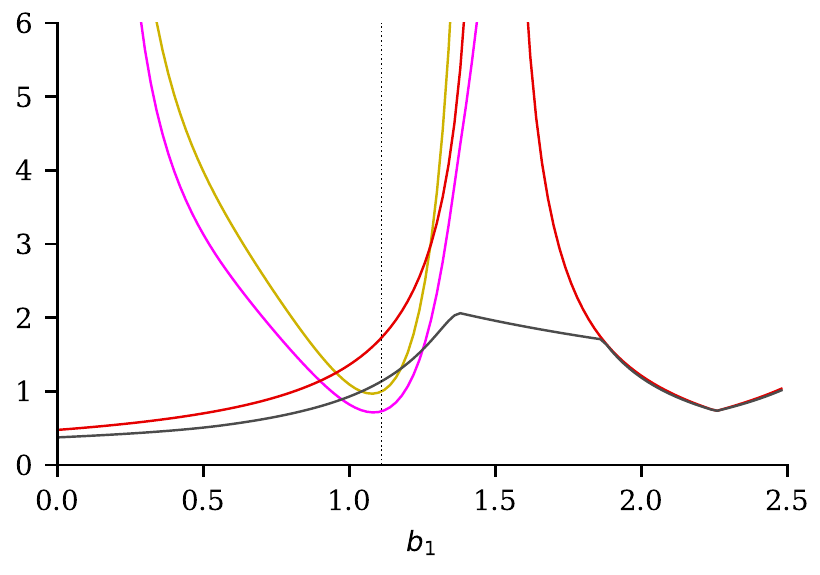}
    \caption{$G$ objective for the standard sextupole harmonics ($P=1$: yellow) and extended harmonics ($P=2$: magenta) as well as  $\max |m|$ ($P=1$: red, $P=2$: dark gray) in dependence of the dipole coefficient $b_1$. The example value of $b_1$ is marked with a dashed line.}
    \label{fig:b1scanA-mG}
\end{figure}

\begin{figure*}[t]
    \centering
    \includegraphics{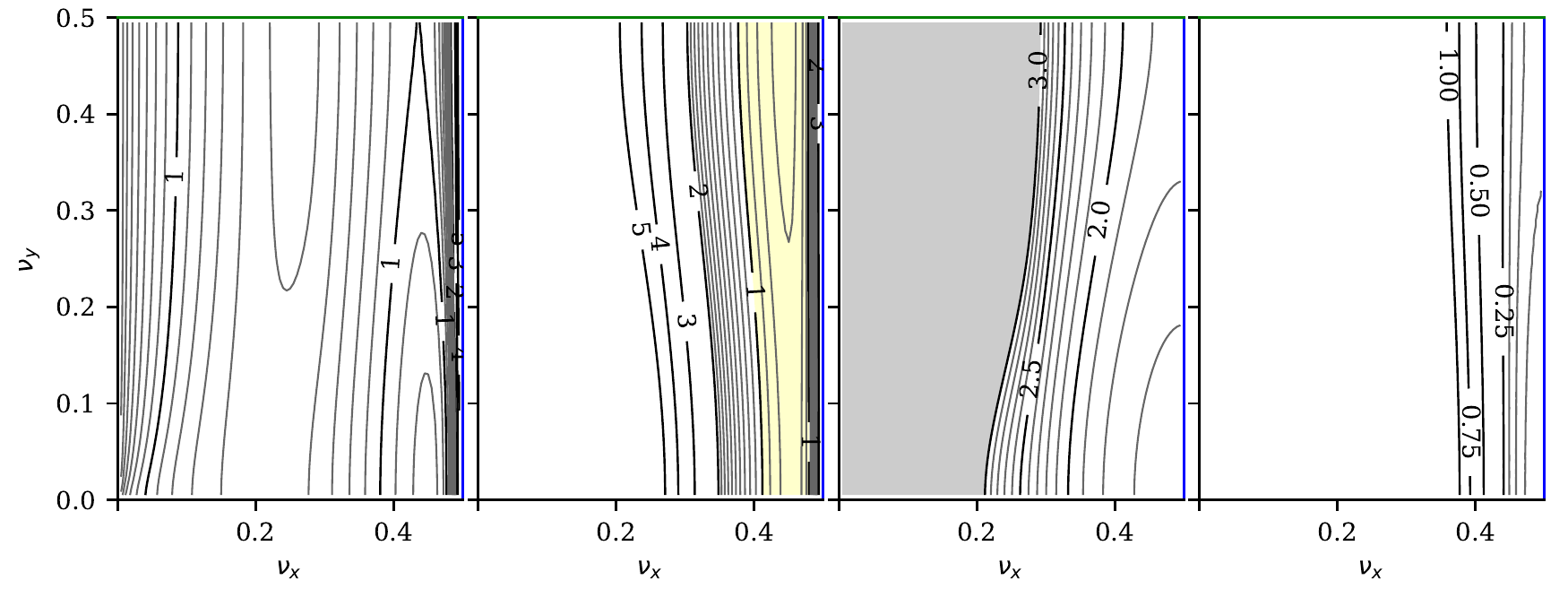}
    \caption{From left to right: (1) optimal $G$ objective, (2) emittance ratio $F$ for $G$ objective, (3--4) quantities like in Fig.~\ref{fig:F} (see legend therein) for optimal $G$.\label{fig:G}}
\end{figure*}

The other region overlaps with the low-emittance regime shown in Fig.~\ref{fig:F}, with the difference that there is now a slight preference for less vertical focusing. The additional parameters $G$ and $\max |m|$ for our example configuration, which is located in that region (see also Table~\ref{tab:example} and Fig.~\ref{fig:b1scanA-int}), are shown in Fig.~\ref{fig:b1scanA-mG}.

\subsection{Extensions to higher harmonics}

Having obtained an optimized solution for the Mathieu cell $(P=1)$, it is possible to iteratively increase $P$ and re-optimize the solution locally. However, the number of free parameters increases significantly. In the scope of this work, we increase to $P=2$ only for the sextupolar field component, so that
\begin{align}
  m(u) &= m_0 + 2 m_1 \cos(2 u) + 2 m_2 \cos(4 u).
\end{align}
This has the advantage that the dimensions of the free parameter space $(\nu_\mathrm x,\nu_\mathrm y,b_1)$ do not increase -- the additional harmonic coefficient $m_2$ is used to reduce $\max |m|$ without changing optical functions.

To compensate chromaticity, we are required to solve a more general variant of Eq.~\eqref{eq:m1chroma}
\begin{align}
  \label{eq:mTriple}
  \vec A_\mathrm x \cdot \vec m = -\xi_\mathrm x, \qquad
  \vec A_\mathrm y \cdot \vec m = \xi_\mathrm y
\end{align}
with the components of $\vec A_\mathrm{x,y}$ holding scaled Fourier components of $\tilde\eta \tilde\beta_\mathrm{x,y}$.
This system is underdetermined; its solution space in three dimensions is given as
\begin{align}
  \vec m = \vec m_{(0)} + a (\vec A_\mathrm x \times \vec A_\mathrm y) \text{ for }a \in \mathbb R,
\end{align}
with $\vec m_{(0)}$ being an arbitrary solution. For our computation we use the least-squares solution of the system \eqref{eq:mTriple}.

The quantity $\max|m|$ can be computed with minor effort, as we require it to be minimal under the constraint of full chromaticity compensation -- this is achieved using an elementary optimization procedure on the scalar $a$.

\vspace{1em}

The results of this optimization in tune space are shown in Fig.~\ref{fig:H}. Relative to the setup using just constant and fundamental harmonic $(P=1)$, an overall reduction of the $G$ objective has been achieved, reaching values $G < 0.7$ in the low-emittance region.

\begin{figure}[!b]
    \centering
    \includegraphics{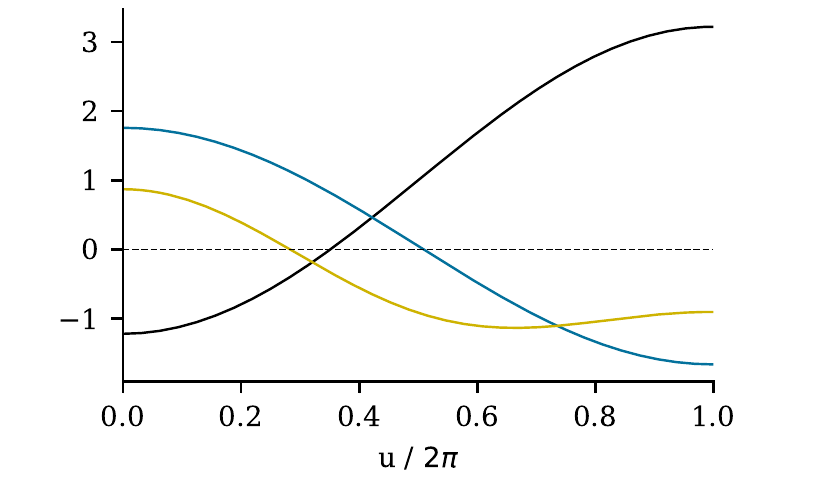}
    \caption{Distribution of dipole (black), quadrupole (blue) and sextupole fields (yellow) for the example solution marked in Fig.~\ref{fig:b1scanA-mG}. The optical functions are identical to those in Fig.~\ref{fig:example}.}
    \label{fig:extendedExample}
\end{figure}

\begin{figure*}
    \includegraphics{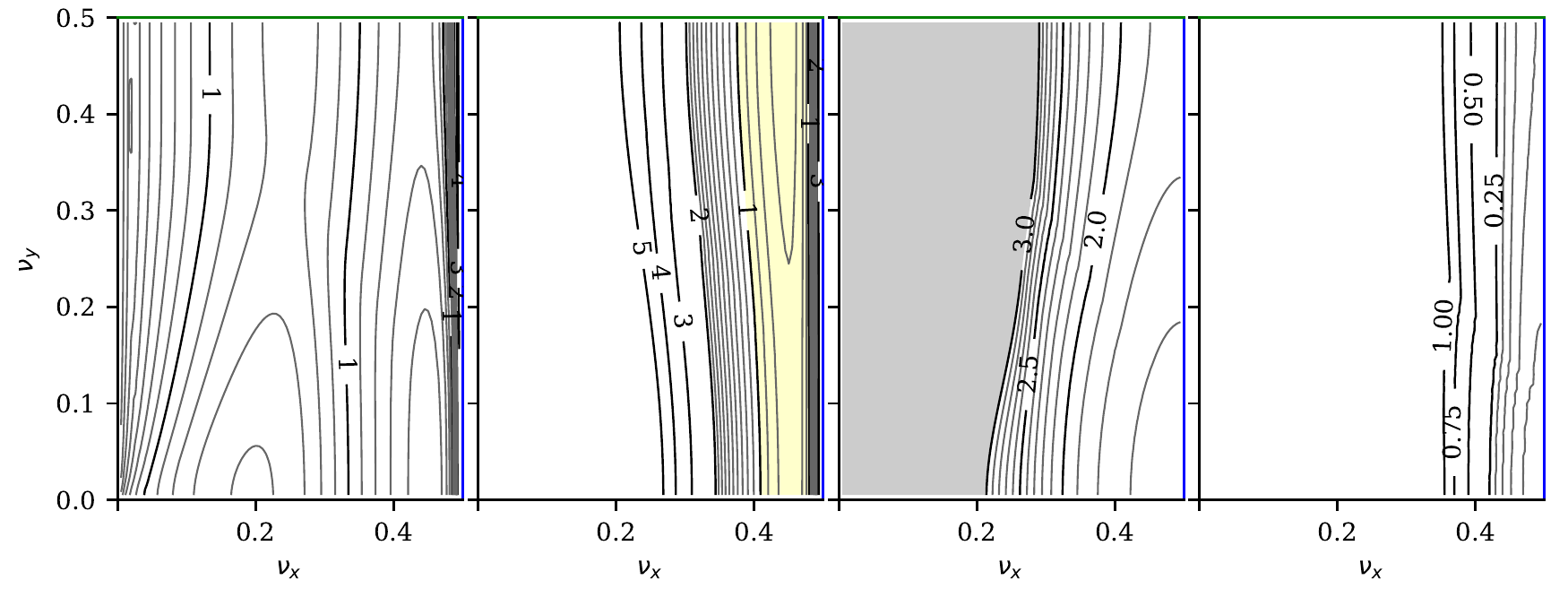}
    \caption{Figures of merit for $G$-optimized solutions with extended sextupole harmonics $(P=2)$ in tune space. See legend in Fig.~\ref{fig:G}.}
    \label{fig:H}
\end{figure*}

This can be observed in more detail for our example tune $\nu_\mathrm x=\num{0.45}$, $\nu_\mathrm y=\num{0.35}$ in Figs.~\ref{fig:b1scanA-mG} and \ref{fig:extendedExample}. The maximum value of $|m(u)|$ has been reduced by decreasing the sextupole strength at the position of maximum bending. This is reasonable as the large sextupolar fields at this location have a negligible influence on chromaticity compensation.

\subsection{Fringe effects\label{sec:fringe}}

The required magnetic potentials can always be constructed in principle, even when including fringe effects. This is discussed in the following in a compressed form, with Appendix~\ref{app:propD} giving more details.

We approximate $(r,\phi,s)$ as a cylindrical coordinate system.
Revisiting Eq.~\eqref{eq:V}, we restrict the basis for $\Psi$ 
\begin{align}
    \Psi &= \sum_{n=1}^3 \sum_{p=0}^2 V_{n,p} \Psi_{n,p},
\end{align}
to match the conditions of symmetry in the $s=0$ transverse plane, as well as symmetry in the $x$-$s$ plane (only upright multipoles) by defining 
\begin{align}
    \Psi_{n,p} &= - D_n(\bar k_p, r) \, \sin(n \phi) \cos(\bar k_p s).
\end{align}
One can obtain the field in the machine plane via the relations
\begin{align}
    \label{eq:ByPhi}
    B_y(x) &\equiv B_\phi(r=x,\phi=0), &\quad
    B_\phi &= -\frac 1r \frac{\tod \Psi}{\tod \phi}.
\end{align}
By series expansion of $B_y(x,s)$ in $x$ (Appendix~\ref{app:propD}), the dipolar and quadrupolar fields on the beam path are given by
\begin{align}
    B_y(0,s) = \sum_p V_{1,p} \cos(\bar k_p s) \\
    \frac{dB_y}{dx}(0,s) = \sum_p V_{2,p} \cos(\bar k_p s).
\end{align}
Note that the linear vertical dependence of $B_y$ is also fixed to be that of a standard quadrupole field by the requirement of vanishing curl in $s$ direction. One can relate the $V$ coefficients to the normalized Mathieu cell coefficients using
\begin{align}
    B_y(0,s) &= (B \rho) \; / \rho(s), &\quad \frac{d B_y}{dx}(0,s) &= (B \rho) \; \kappa(s).
\end{align}

For the sextupolar fields, the relation is given by Eq.~\eqref{eq:defMu}. Here, mixing with the $V_{1,1}$ components from the dipole potential occurs, as
(Appendix~\ref{app:propD})
\begin{align}
    \frac{d^2 B_y}{d x^2}(0,s) = V_{3,0} + V_{3,1} (1+f) \cos(\bar k_1 s) + V_{3,2} \cos(\bar k_2 s),    
\end{align}
with the fringe factor
\begin{align}
    f = -\left( \frac{\theta}{\pi} \right)^2 \frac{b_1}{2 \bar m_1}.
\end{align}
Here, $\bar m_1$ is the value of the $m_1$ sextupole coefficient when ignoring fringe effects. With fringe effects, the value of $m_1$ is shifted to \ref{app:propD}
\begin{align}
    m_1 = \bar m_1 (1 + f).
\end{align}
As an estimate for the typical strength of the fringe effect, assuming $|b_1|, |m_1| \sim 1$, we can use the expression $(\theta / \pi)^2$ occuring in $f$. 
For bending angles per cell of $\theta \leq \SI 5{deg}$, we obtain $(\theta / \pi)^2 \leq 1/36^2 < \num{e-3}$.

For larger values of $|f|$, the sextupole strength can always be readjusted to yield the proper value of $m_1$. Note that with the fringe effect, $m_1$ does not refer to a standard sextupole in the transverse plane anymore -- however, we only require the field in the machine plane to compensate chromaticity via horizontal dispersion; the quadrupolar fields in the machine plane are always properly defined due to the condition of vanishing curl as stated.

\section{SLS 2.0 example\label{sec:sls2}}

The Swiss Light Source upgrade (SLS 2.0) has a unit cell length of \SI{2.165}m and a unit cell bending angle of \SI 5{deg}. The average curvature radius and the characteristic magnetic field density from Eq.~\eqref{eq:charB} are approximated using these values as
\begin{align}
    1 / \avg{1 / \rho} &= \SI{24.81}m, & B_\text c &= \SI{322.5}{mT}.
\end{align}
According to Eq.~\eqref{eq:maxB} and assuming a normal-conducting magnet limit of $\max B \sim \SI 2T$, we get $\max|b| \sim 6.2$, or $\max |b_1| \sim 2.6$.

We assume the maximum applicable sextupole strength at $\max |\mu| = \SI{650}{m^{-3}}$, which is a conservative estimate consistent with the present lattice design. 
By using Eq.~\eqref{eq:opt-length} we are able to compute the optimal cell length for a Mathieu cell with example parameters for SLS 2.0. 
Using the standard sextupole harmonics ($P=1$, Table~\ref{tab:example}) we obtain $\max |m| \sim 1.726$, resulting in an optimal cell length of $\sim\SI{1.592}m$. Using the extended sextupole harmonics ($P=2$, Fig.~\ref{fig:b1scanA-mG}) we obtain a reduced value of $\max |m| \sim 1.134$, resulting in an optimal cell length of $\sim \SI{1.433}m$.

\subsection{Improved optimal cell length estimate using pole-tip fields}

It should be noted that, due to the overlapping of fields with different multipolar order, the pole tip field of a combined-function magnet will be higher than that of the sextupole component, thus increasing the optimal cell length. For a detailed example we calculate pole-tip fields $B_r^\text{pt}$ with the common approach~\cite{opa}, i.e., without considering longitudinal variation as in Eq.~\eqref{eq:V}, as
\begin{align}
   &\frac{B_r^\text{pt}(s,\phi,R)}{(B \rho)} = \frac{\sin \phi}{\rho(s)} + \sin(2 \phi) \kappa(s) R + \sin(3\phi) \mu(s) R^2
\end{align}
with the pole-tip radius $R$, or as a unitless equation,
\begin{equation}
\label{eq:BrBc}
\begin{aligned}
  \frac{B_r^\text{pt}}{B_\text c}(u,\phi,R) = &\sin \phi \, b(u) \\ +&\sin(2 \phi) k(u) \left(\frac{L_\text{c}}L\right)^2 \\
  +&\sin(3\phi) m(u) \left(\frac{L_\text{c}}L\right)^4
\end{aligned}
\end{equation}
where we defined the characteristic length
\begin{align}
    L_\text c = \pi \sqrt{ R / \avg{ 1 / \rho} }
\end{align}
containing the geometric mean of chamber and average curvature radius.
In the case of interest, the maximum pole-tip field strength is not dominated by $m(s)$ alone, as would be the case for $L_\text{c} / L \gg 1$. Instead, the situation $L_\text{c} \sim L$ occurs because multipoles of different order often have comparable pole-tip field magnitudes.

While the pole-tip field can be used as an estimate for the technical feasibility of magnet design, this estimate can be improved further. To do so, we take into account the empirical knowledge that the feasible pole-tip fields decrease with the multipole order $n$ -- e.g., for the SLS 2.0 separate-function magnets we may assume an inverse relation $\max B_r^\text{pt} \sim \SI 2T / n$.

To include the improved estimate for combined-function magnets, their contributions from Eq.~\eqref{eq:BrBc} are weighted with their order, leading to the definition of a weighted pole-tip field via
\begin{equation}
\begin{aligned}
  \frac{B_r^\text{w}}{B_\text c}(u,\phi,R) = &\phantom1 \sin \phi \, b(u) \\ +&2 \sin(2 \phi) k(u) \left(\frac{L_\text{c}}L\right)^2 \\
  +&3 \sin(3\phi) m(u) \left(\frac{L_\text{c}}L\right)^4
\end{aligned}
\end{equation}

\begin{figure}[!b]
    \includegraphics{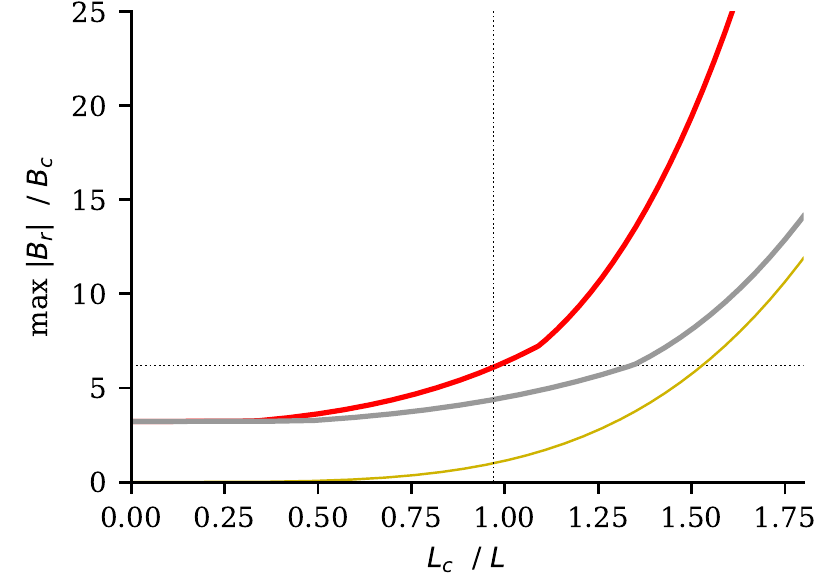}
    \caption{Maximum weighted pole-tip field $B_r^\text w$ (red), actual pole-tip field $B_r^\text{pt}$ (gray), and pole-tip field of sextupole component (yellow) in dependence of normalized inverse cell length $L_\text c / L$, all in units of characteristic field density and length. The values of $\max|B_r|=\SI 2T$ for SLS 2.0 assumptions and the corresponding value of $L_\text c / L$ for a maximum weighted pole-tip field close to that strength are denoted by dotted lines.\label{fig:poleTip}}
\end{figure}

We can obtain good approximations of the maximum pole-tip fields for a given value of $L_\text c / L$ by computing the maximum value of $B_r / B_\text c$ on a grid of $(\phi, u)$ points. In this work we use 128 values of $u$ and 16 values of $\phi$. The result of this computation with the example cell is shown in Fig.~\ref{fig:poleTip}. One can observe that, as expected, the sextupole strengths dominate for large $L_\text c / L$; small values are dominated by the constant dipole contribution.

\subsection{SLS 2.0 parameters and results}

For SLS 2.0 we assume a chamber radius $R = \SI{10}{mm}$ and obtain the characteristic length $L_\text{c} \sim \SI{1.565}m$. The technical limit of pole-tip fields in such a distributed magnet structure is yet to be determined. Comparing the actual pole-tip field in Fig.~\ref{fig:poleTip} with the sextupole-only contribution, we can see that the optimal cell length increases significantly when all multipoles are considered.

\begin{figure}
    \includegraphics{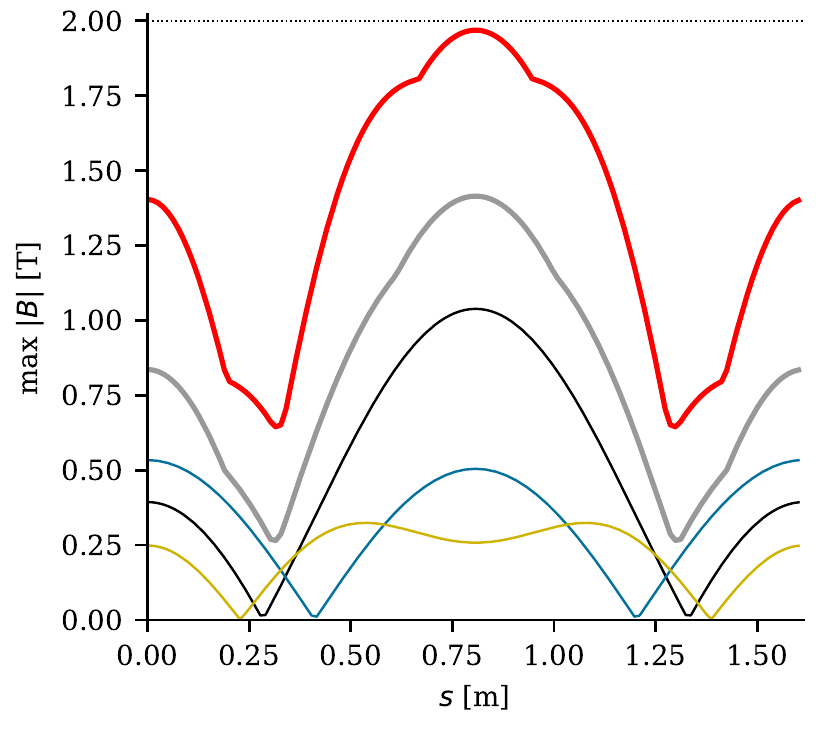}
    \caption{Absolute maxima of dipole (black), quadrupole (blue) and sextupole (yellow) pole-tip fields, actual pole-tip field $B_r^\text{pt}$ (gray), and weighted pole-tip field $B_r^\text w$ (red) along $s$ for the SLS 2.0 example (cf.~Fig.~\ref{fig:opa-lattice}).}
    \label{fig:poleTipContribs}
\end{figure}

We now consider the example values marked in  Fig.~\ref{fig:poleTip},  where the optimal cell length is $L \sim \num{1.031} L_\mathrm c \sim \SI{1.614}m$ and $\max|B_r^\mathrm w|$ is close to \SI 2T with a small safety margin. The distribution of multipole contributions to the pole-tip fields is shown in Fig.~\ref{fig:poleTipContribs}. 

This example magnet configuration is analysed using the optics code \textsc{opa}~\cite{opa}. As optics codes usually do not work in Fourier space, we discretize the solution into segments of dipole-quadrupoles and thin sextupoles. For convenience, we choose 128 segments for each magnet type.

\begin{figure*}
    \subfloat[Pole-tip fields for multipole slices (cf.~Fig.~\ref{fig:poleTipContribs}). Only fields for multipoles of finite length (dipole: blue, quadrupole: red) are shown.]
    {\includegraphics[width=0.49\textwidth]{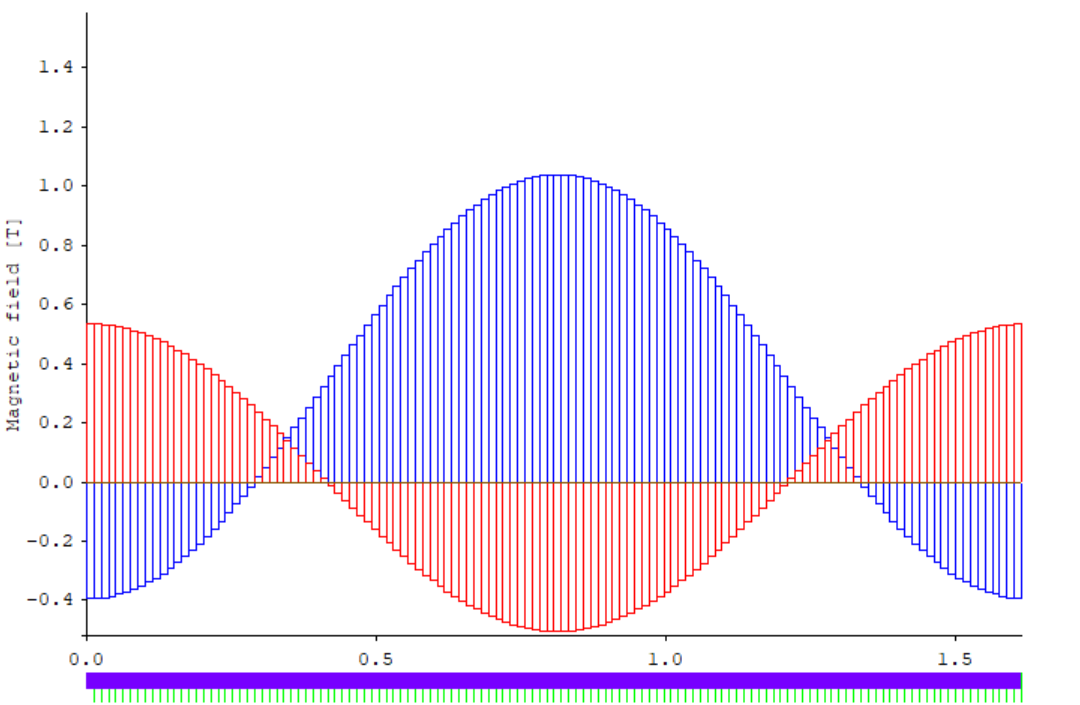}}
    \subfloat[optical functions of optimized Mathieu cell for SLS 2.0 in \textsc{opa} ($\beta_x$: blue, $\beta_y$: red, $\eta$: green)]
    {\includegraphics[width=0.49\textwidth]{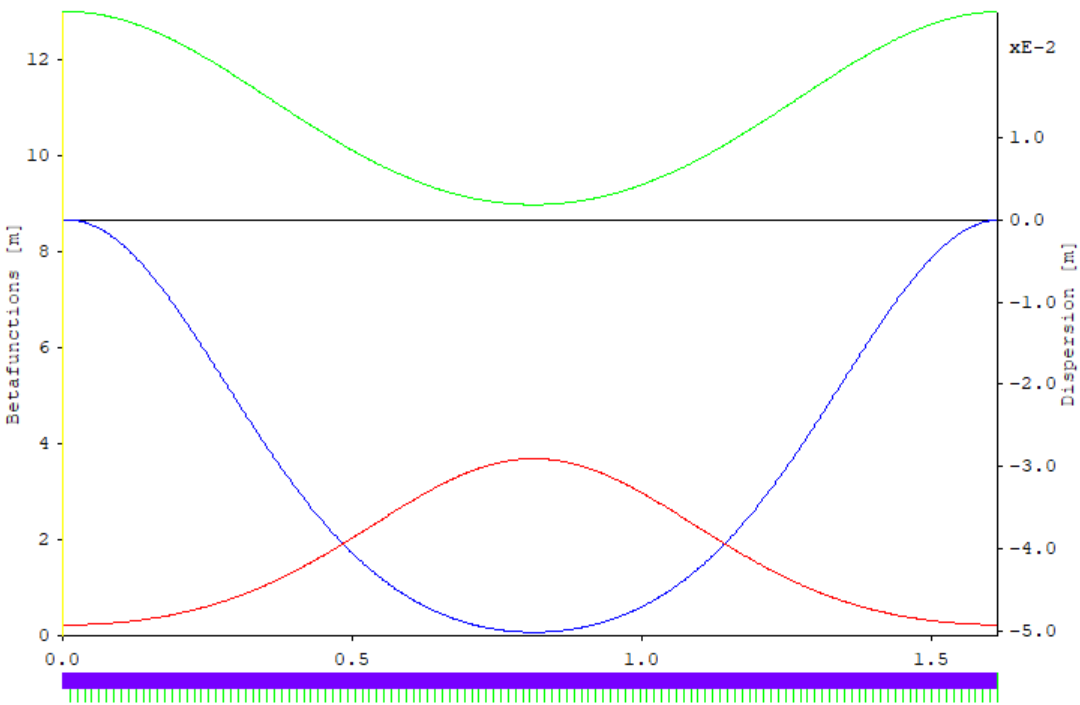}}
    \caption{Properties of the Mathieu cell for the SLS 2.0 example as computed by \textsc{opa}.}
    \label{fig:opa-lattice}
\end{figure*}

The optics results are shown in Fig.~\ref{fig:opa-lattice}, and Table \ref{tab:opa} shows global figures of merit as computed by \textsc{opa}. For the betatron tunes, we can observe that for our example, neglecting weak focusing and edge focusing as stated in Sec.~\ref{sec:bending} is justified. Within the assumptions about pole-tip fields, which may exceed technical limits, and our assumptions about weak focusing and general feasibility of the non-trivial magnetic field arrangement, we obtain an emittance of $\sim \SI{33.2}{pm}$, which is significantly less than the SLS 2.0 design of $\sim \SI{100}{pm}$ \cite{andreas-jsr}. The Mathieu cell's horizontal damping partition is in the range of SLS 2.0 designs with the present $J_x$ also being $\sim 1.8$ in difference to earlier designs \cite{andreas-jsr}. The $\sim \SI{760}{keV}$ energy loss per turn is slightly more than the present SLS 2.0 design at $\sim \SI{690}{keV}$, while the damping times of both the Mathieu cell and the SLS 2.0 unit cell are both in the order of a few \si{ms}.

In addition to the aforementioned complications, the cell is almost isochronous with a momentum compaction in the \num{e-6} range. This can be circumvented by a minor decrease of $b_1$ at the expense of slightly increased emittance (see Fig.~\ref{fig:b1scanA-int}).

\begin{table}[htbp]
    \caption{Lattice parameters of the Mathieu cell for the SLS 2.0 example as computed by \textsc{opa}. Rounded values have been used where appropriate.\label{tab:opa}}
    \begin{ruledtabular}
    \begin{tabular}{rr}
    \textbf{Parameter} & \textbf{Value} \\ \hline
cell length          &  \SI{1.6140}m \\
horizontal cell tune $\nu_\mathrm x$  &  \num{0.45057} \\
vertical cell tune $\nu_\mathrm y$  &  \num{0.34988} \\
nat. horizontal cell chromaticity $\xi_\mathrm x$ &\num{-1.84995} \\
nat. vertical cell chromaticity $\xi_\mathrm y$ &\num{-0.67107} \\
momentum compaction  &  \num{-2.718E-06} \\
horizontal damping partition $J_\mathrm x$ & \num{1.8207} \\ \hline
beam energy                  &  \SI{2.4}{GeV} \\
radiated energy / cell passage &  \SI{4.243}{keV} \\
natural energy spread &  \num{8.6422E-04} \\
horizontal damping time   & \SI{3.345}{ms} \\
vertical damping time    & \SI{6.090}{ms} \\
longitudinal damping time    & \SI{5.164}{ms} \\
horizontal emittance $\epsilon_\mathrm x$ &  \SI{33.19}{pm} \\ \hline
$I_1$ integrated over cell &  \SI{-4.387E-06}m \\
$I_2$ integrated over cell &  \SI{9.084E-03}{m.^{-1}} \\
$I_3$ integrated over cell &  \SI{9.467E-04}{m.^{-2}} \\
$I_4$ integrated over cell &  \SI{-7.455E-03}{m.^{-1}} \\
$I_5$ integrated over cell &  \SI{6.495E-08}{m.^{-1}} \\
    \end{tabular}
    \end{ruledtabular}
\end{table}

\section{Conclusion}

In this work, we introduced Mathieu unit cells as elementary approximations for periodic lattice systems.
Due to their distributed multipolar structure, they allow for the inclusion of combined-function effects, as well as the computation of common figures of merit like momentum compaction and emittance. They even predict the usefulness of combining longitudinal gradients with reverse bending, and reach sub-TME emittance.

By its nature, the goal of this study can only be to illuminate an 'undercurrent' of the sinusoidal focusing concept, permeating lattice design, that otherwise is concealed by practical requirements of accelerator technology. It goes without saying that a realizable lattice design requires detailed studies incorporating a multitude of boundary conditions \cite{emma-raubenheimer, icfa-57} e.g., dynamic aperture considerations, robustness in case of field deviations \cite{johan-ler14,johan-sls2rep}, which depend on the cell length in a non-trivial manner. 

Mathieu cells are useful tools for investigating basic lattice configurations and performance limits. Sinusoidal bending forces are commonly used in the description of undulator fields (see e.g.~\cite{wille}). In the context of further progress on MBA miniaturization and combined-function magnet lattices, e.g., \cite{bogomyagkov,tavares-miniba,yang-bai,complexbend}, the concept of Mathieu cells could thus help to shape future lattice designs.

The source code for all computations in this work, excluding the ones performed in \textsc{opa}, is based on the SciPy framework~\cite{scipy,matplotlib} and fully accessible~\cite{mathieu-cell}.

\begin{acknowledgments}
The author thanks M.~Kranj\v{c}evi\'{c}, J.~Kallestrup, A.~Streun and J.~Bengtsson for improving the manuscript by proofreading and/or hinting at useful references. Furthermore, the author appreciates the general support of M.~Aiba, M.~B\"oge, J.~Chrin, and T.~Schietinger.
\end{acknowledgments}

\appendix
\section{scaling cell length in a fixed arc\label{app:fxArc}}

To obtain results as general as possible, this work often uses multipoles normalized to a dimensionless unit cell. A rule is that the standard, rigidity-normalized multipole fields -- $1/\rho$ for curvature, $\kappa$ for quadrupole focusing strength, $\mu$ for sextupole strength -- are denoted by Greek letters. Their cell-normalized, dimensionless counterparts -- $b$ for normalized curvature, $k$ for normalized quadrupole, $m$ for normalized sextupole strength, are denoted by Latin letters.

The dimensionless, cell-normalized optics functions are also denoted with a tilda, $\tilde \bullet$, to distinguish them from the standard optics functions. All occuring synchrotron integrals $I_\bullet$ are cell-normalized.

When replacing the path length $s$ by a scaled path length $u=\pi s / L$, we require the scaled solution $x(u)$ to fulfill Hill's equation \eqref{eq:hill}
\begin{align}
  \frac{\tod^2}{\tod s^2} x(u) + \kappa(u) x(u) &= 0,
\end{align}
so that
\begin{align}
  \frac{\tod^2}{\tod u^2} x(u) + [ (L / \pi)^2 \kappa(u) ] \; x(u) &= 0
\end{align}
and by comparison
\begin{align}
  \kappa(u) = \left( \frac{\pi}L \right)^2 k(u),
\end{align}
resulting in the standard quadrupole strength scaling with the inverse square of cell length.

Since we require the tune for all cells to be independent of the cell length, this should also apply to the natural chromaticity so that $\int \beta \kappa \,\tod s \propto \beta / L$ is constant, and 
\begin{align}
  \beta(u) = \frac L{\pi} \tilde\beta(u)
\end{align}
is linear in $L$, so $\int \tilde\beta k\,\tod s$ is also constant, with $\tilde\beta$ being the cell-normalized optics function.

Furthermore, the dispersion function $\eta(s)$ must fulfill the inhomogeneous Hill's equation
\begin{align}
  1/\rho(u) &= \frac{\tod^2}{\tod s^2} \eta(u) + \kappa(u) \eta(u).
\end{align}
As the average arc curvature should remain constant, we require $b(u)$ to be independent of cell length. Division by $\avg{1/\rho}$ yields
\begin{equation}
\begin{aligned}
  b(u) &= \frac{\tod^2}{\tod s^2} \frac{\eta(u)}{\avg{1/\rho}} + \kappa(u) \frac{\eta(u)}{\avg{1/\rho}}\\
      &= \frac{\tod^2}{\tod s^2} \tilde \eta(u) + k(u) \tilde \eta(u)
\end{aligned}
\end{equation}
with
\begin{align}
  \tilde\eta(u) &= \left( \frac{\pi}L \right)^2 \frac{\eta(u)}{\avg{1/\rho}}.
\end{align}
For the compensated chromaticity to be independent of cell length, we require 
\begin{equation}
\begin{aligned}
    &\null \int \beta(u) \eta(u) \mu(u) \tod s\\
    &= \left( \frac L{\pi} \right)^4 \avg{1/\rho} \int \tilde\beta(u) \tilde\eta(u) \mu(u) = \text{const.} 
\end{aligned}
\end{equation}
Then sextupole strength scales as
\begin{align}
    \label{eq:muScale}
    \mu(u) = \left( \frac{\pi}L \right)^4 \frac{m(u)}{\avg{1/\rho}}. 
\end{align}
Note that this inverse quartic scaling is due to the average curvature $\avg{1 / \rho}$ remaining constant -- if the ring was miniaturized as a whole, $\avg{1 / \rho} \propto 1/L$ would hold, resulting in inverse cubic scaling and corresponding to the multipole order.

\section{Properties of scaled Bessel function\label{app:propD}}

Using the series expansion of $I_n$ \cite{abramovitz-stegun}
\begin{align}
  I_n(x) = (x/2)^n \sum_{q=0}^\infty \frac{(x/2)^{2 q}}{q! (n+q)!}
\end{align}
and the definition
\begin{align}
  D_n(k, x) &= I_n(k x) \Big/ (k/2)^n,
\end{align}
one obtains
\begin{align}
    \label{eq:DnSeries}
    D_n(k,x) = x^n \sum_{q=0}^\infty (k/2)^{2 q} \frac{ x^{2 q}}{q! (n+q)!}.
\end{align}
Note that a removable singularity exists at $k=0$, 
\begin{align}
    \lim_{k\rightarrow 0} D_n(k,x) = x^n / n!.
\end{align}

The derivative, required for the radial component of field density, can be expressed as a series in $k$,
\begin{align}
    \frac{\tod D_n(k,x)}{\tod x} &= x^{n-1} \sum_{q=0}^\infty (2q+n) \frac{x^{2 q}}{q!(n+q)!} \nonumber \\
    &= n \frac{x^{n-1}}{n!} + (2+n) \left( \frac k2 \right)^2 \frac{x^{n+1}}{(n+1)!} + \dots \nonumber \\
    &= \frac{x^{n-1}}{(n-1)!} \left[ 1 + k^2 \frac{n+2}{4 n (n+1)} x^2 + \dots \right].
\end{align}

We are also interested in the field density in the machine plane, given by Eq.~\eqref{eq:ByPhi}
\begin{align}
    \label{eq:ByD}
    B_y(x,s) = \sum_{n=1}^3 \sum_{p=0}^2 V_{n,p} \frac{D_n(\bar k_p, x)}x n \cos(\bar k_p s),
\end{align}
specifically its series expansion in $x$, yielding the multipole components. Inserting Eq.~\eqref{eq:DnSeries} into
Eq.~\eqref{eq:ByD}, one obtains an approximation for small $x$ as
\begin{align}
    B_y(x,s) &= \sum_{p=0}^2 \cos(\bar k_p s) \cdot \nonumber \\ &\cdot \left[ V_{1,p} + V_{2,p} x + (V_{3,p} + V_{1,p} \bar k_p^2 / 4) \frac{x^2}2 + O(x^3) \right].
\end{align}
Utilizing that $\bar k_0=0$, and that the dipolar and quadrupolar fields are set to zero for the $p=2$ harmonic ($V_{3,1}=V_{3,2}=0$), we can simplify this series to
\begin{widetext}
\begin{align}
    B_y(x,s) = &V_{1,0} + V_{1,1} \cos(\bar k_1 s) + \Big[ V_{2,0} + V_{2,1} \cos(\bar k_1 s) \Big] x + \nonumber \\
    &+ \Big[ V_{3,0} + \left(V_{3,1} + V_{1,1} \frac{\bar k_1^2}4 \right) \cos(\bar k_1 s) + V_{3,2} \cos(\bar k_2 s) \Big] \frac{x^2}2 + O(x ^3).
\end{align}
\end{widetext}
For almost all coefficients $V$, we find that each is proportional to a dipolar, quadrupolar, or sextupolar focusing term. The only exception is the quadratic coefficient
\begin{align}
   \left(V_{3,1} + V_{1,1} \frac{\bar k_1^2}4 \right),
\end{align}
feeding up the dipolar fringe effect via the $V_{1,1}$ coefficient. To quantify its effect, we define a fringe factor
\begin{align}
    f = \frac{\bar k_1^2}4 \frac{V_{1,1}}{V_{3,1}} = \frac{\pi^2}{L^2} \frac{V_{1,1}}{V_{3,1}},
\end{align}
so that
\begin{align}
    \frac{d^2 B_y}{d x^2}(0,s) = V_{3,0} + V_{3,1} (1+f) \cos(\bar k_1 s) + V_{3,2} \cos(\bar k_2 s).
\end{align}
And, by using Eq.~\eqref{eq:defMu} and Eq.~\eqref{eq:muScale},
\begin{align}
    m(u) = m_0 + 2 \bar m_1 (1+f) \cos(2u) 2 + \bar m_2 \cos(4u).
\end{align}
with $\bar m_1$ being the sextupole coefficient set when ignoring the fringe effect ($f=0$), and
\begin{align}
    m_1 = \bar m_1 (1 + f).
\end{align}
Again by connecting the $V$ coefficients in $f$ to the $b$ and $m$ components via 
 Eq.~\eqref{eq:defMu} and Eq.~\eqref{eq:muScale}, one obtains
\begin{align}
  f = -\left( \frac{L \avg{1/\rho}}{\pi} \right)^2 \frac{b_1}{\bar m_1} = -\left( \frac{\theta}{\pi} \right)^2 \frac{b_1}{2 \bar m_1}.  
\end{align}

\bibliography{refs}

\end{document}